\documentclass[aps,preprint,nofootinbib]{revtex4}%
\usepackage{amsmath}
\usepackage{graphicx}
\usepackage{amsfonts}
\usepackage{epsfig}
\usepackage{amssymb}%
\usepackage[usenames]{color}

\providecommand{\U}[1]{\protect\rule{.1in}{.1in}}
\providecommand{\U}[1]{\protect\rule{.1in}{.1in}}

\begin{document}
\title{A new approach to the vacuum of inflationary models}
\author{Shih-Hung Chen}
\email{schen102@asu.edu}
\affiliation{Department of Physics and School of Earth and Space Exploration, Arizona State
University, Tempe, AZ 85287-1404}
\author{James~B.~Dent}
\email{jbdent@asu.edu}
\affiliation{Department of Physics and School of Earth and Space Exploration, Arizona State
University, Tempe, AZ 85287-1404}

\begin{abstract}
A new approach is given for the implementation of boundary conditions used in
solving the Mukhanov-Sasaki equation in the context of inflation. The familiar
quantization procedure is reviewed, along with a discussion of where one might
expect deviations from the standard approach to arise. The proposed method
introduces a (model dependent) fitting function for the $z^{\prime\prime}/z$
and $a^{\prime\prime}/a$ terms in the Mukhanov-Sasaki equation for scalar and
tensor modes, as well as imposes the boundary conditions at a finite conformal
time. As an example, we employ a fitting function, and compute the spectral
index, along with its running, for a specific inflationary model which
possesses background equations that are analytically solvable. The
observational upper bound on the tensor to scalar ratio is used to constrain
the parameters of the boundary conditions in the tensor sector as well. An
overview on the generalization of this method is also discussed.

\end{abstract}
\maketitle



\newpage

\section{Introduction\label{intro}}

As is well known, the inflationary paradigm provides solutions to many
cosmological problems such as the flatness problem, the horizon or causality
problem, and also dilutes unwanted (and unobserved) relics \cite{Guth, Linde,
Steinhardt}. It also provides a natural mechanism of producing primordial
perturbations that seed the inhomogeneities of the universe \cite{Mukhanov
pert,sasaki pert}. The basic idea is that the quantum fluctuations of a
classically homogeneous scalar field, the inflaton, source quantum
fluctuations of the spacetime metric (the inflaton will create density
perturbations which will source the scalar fluctuations of the metric). During
the process of inflation, this quantum fluctuation is amplified to become a
classical fluctuation, and, at the end of inflation, the fluctuation in the
metric induces the density fluctuations of matter that were produced during
reheating . This primordial perturbation generated during inflation then is
what gives rise to the formation of structure in the universe.

In this story, the crucial quantities to be determined are the amplitude of
the primordial density and tensor perturbations, as the growth of structure is
dependent on their size. The subsequent evolution of the primordial
perturbations can be inferred from careful observations of the history of the
growth of structure. From this we expect that the density perturbation
produced by inflation to be $\mathcal{O}(10^{-5})$.

Therefore, it is important to be able to accurately compute these
perturbations in order to either preserve or rule out a given inflationary
model. For the usual models of inflation, Ricci scalar plus a single
canonically normalized scalar field, there are two components that will
determine this amplitude. As will be described in more detail below, the first
is the form of a function, $z^{\prime\prime}/z$, which arises in the
Mukhanov-Sasaki equation. This equation is satisfied by a mode function,
$v_{k}$, which arises by a redefinition of the co-moving curvature
perturbation in momentum space, $\mathcal{R}_{k}$, upon having written the
original action in terms of $\mathcal{R}_{k}$. Since this equation is crucial
to finding the curvature perturbation (the same equation is also obeyed by the
tensor modes), it is essential that one accurately specifies its form.

The second component is the input from vacuum selection, which is equivalent
to a boundary condition for the Mukhanov-Sasaki equation. For example, one
avenue of study has been to alter the initial state to lie away from the
standard Bunch-Davies vacuum\cite{BunchDavies,Danielsson trans Planck,New
Physics CMB,Easther:2005a,Chen:2006a,Holman:2007a,Meerburg:2009a,Padmanabhan,Ashoorioon:2012}.
Such alternative boundary conditions are typically chosen by conditions set at
a given cut-off scale in either momentum or time. These choices will then
manifest themselves in physical observables (for example as new features in
the power spectrum, or enhanced non-Gaussianity) which can allow one to gain
knowledge of the initial state from observation.  It should be mentioned that there exist arguments \cite{Susskind} that the Bunch-Davies vacuum might be the most
probable vacuum to produce the correct power spectrum from the perspective of technical naturalness, although this does not eliminate the possibility of deviations from the Bunch-Davies vacuum.

It is customary to characterize inflation as a period of time where the scale
factor grew almost exponentially, a period called de Sitter or quasi-de Sitter
inflation (exact exponential growth of the scale factor, $a\propto e^{Ht}$,
with $H$ constant, is technically de Sitter inflation, and nearly exponential
growth is termed quasi-de Sitter). If the universe inflates as a power law
manner $a\varpropto\tau^{p}$ where $\tau$ is the conformal time and $p$ is a
constant, than the solution of the Mukahanov-Sasaki equation is known. Notice
that de Sitter inflation correspond to $p=-1.$ The solution for general $p$ is
given by a linear combination of Bessel functions. The calculation for the
perturbation amplitude has been well established for such a case.\cite{Lyth
and Stewart} However, for most models of inflation, power law expansion
happens only in a short period of time either at the beginning or the end of
the inflation. Thus this commonly used approximation may not apply to all
inflationary models. If one insists on using the equations derived from the
power law limit, one runs the risk of possibly ruling out phenomenologically
viable models, or of preserving models that are ruled out by observational data.

In the present work we would like to address the possibility that the function
$z^{\prime\prime}/z$ inside the Mukhanov equation deviates from the de Sitter
limit, and how that may affect one's choice of boundary conditions. It may be
that before some point $\tau_{p}$ that using the de Sitter limit is not
consistent, and therefore placing boundary conditions at $\tau_{p}$ is more
natural. We have thus expanded the standard method of computing the amplitude
of the primordial perturbation for such a case. Our method applies to those
inflationary models that do not behave with power law (at least partially)
expansion with some specific constraints in the background evolution.

The paper is organized as follows. In Sec.\ref{standardmethod} we review the
standard calculation of primordial density and tensor perturbations where the
de Sitter limit is taken. We also explain the physical reasons behind the
commonly chosen Bunch-Davies vacuum. In Sec.\ref{vacuumselection} we introduce
a new method of vacuum selection by applying this method to a specific
inflation model. The principles of generalizing this method to other models is
also given. In Sec.\ref{conclusions} we give our conclusions.

\section{The standard method}

\label{standardmethod}

We will now outline the key ingredients for the calculation of the primordial
perturbations by quantizing the comoving curvature perturbation as well as the
tensor perturbation \cite{Lyth and Stewart} (for a recent textbook treatment
and lecture notes see \cite{Weinberg:2008}). The theories we consider here
contain an Einstein-Hilbert action and a canonically normalized scalar field,
minimally coupled to gravity with an arbitrary self-interacting potential
\begin{equation}
S=\int d^{4}x\sqrt{-g}\left\{  \frac{1}{2\kappa^{2}}R-\frac{1}{2}g^{\mu\nu
}\partial_{\mu}\sigma\partial_{\nu}\sigma-V\left(  \sigma\right)  \right\}
\label{action}%
\end{equation}
where $\kappa^{2}=8\pi G=1/M_{Pl}^{2}$ and our metric signature is $-+++$.

\subsection{Scalar Perturbations}

We begin with the perturbed Friedmann-Robertson-Walker(FRW) metric including
the most general perturbations%
\begin{equation}
ds^{2}=a^{2}(\tau)\{-(1+2A)d\tau^{2}-2\partial_{i}Bdx^{i}d\tau+\left[  \left(
1+2\mathcal{R}\right)  \delta_{ij}+\partial_{i}\partial_{j}H_{T}\right]
dx^{i}dx^{j}\}
\end{equation}
Where $A\left(  \tau,\mathbf{x}\right)  ,B\left(  \tau,\mathbf{x}\right)
,\mathcal{R}\left(  \tau,\mathbf{x}\right)  $ and $H_{T}$ $\left(
\tau,\mathbf{x}\right)  $ are small perturbations around homogeneous FRW
metric. We will be concerned with calculating the scalar $\mathcal{R}$, which
is the gauge invariant comoving curvature perturbation.

Variation of the action Eq.(\ref{action}) gives the Einstein equations and the
scalar field equation of motion, which at the background level are%
\begin{align}
&  \frac{\left(  a^{\prime}\right)  ^{2}}{a^{4}}=\frac{\kappa^{2}}{3}\left[
\frac{1}{2a^{2}}\left(  \sigma^{\prime}\right)  ^{2}+V\left(  \sigma\right)
\right] \label{00}\\
&  \frac{a^{\prime\prime}}{a^{3}}-\frac{\left(  a^{\prime}\right)  ^{2}}%
{a^{4}}=-\frac{\kappa^{2}}{3}\left[  \frac{1}{a^{2}}\left(  \sigma^{\prime
}\right)  ^{2}-V\left(  \sigma\right)  \right] \label{11}\\
&  \frac{\sigma^{\prime\prime}}{a^{2}}+2\frac{a^{\prime}}{a^{3}}\sigma
^{\prime}+V^{\prime}\left(  \sigma\right)  =0 \label{eom sigma}%
\end{align}
where a prime denotes a derivative with respect to the conformal time, $\tau$,
while a dot will indicate a derivative with respect to the coordinate time,
$t$.

Putting the solutions for the background evolution back into the
Einstein-Hilbert action, and expanding the action to second order in the
perturbations gives (setting $\kappa^{2}=1$)
\begin{equation}
S_{(2)}=\frac{1}{2}\int d^{4}xa^{3}\frac{\dot{\sigma}^{2}}{H^{2}}\left[
\mathcal{\dot{R}}^{2}-a^{-2}\left(  \partial_{i}\mathcal{R}\right)
^{2}\right]  .
\end{equation}
The above expression can be obtained using the gauge symmetry in the action to
choose $\delta\sigma=0.$ One may define the Mukhanov variable
\begin{equation}
v\equiv z\mathcal{R},\text{ \ \ where \ \ \ }z^{2}\equiv a^{2}\frac
{\dot{\sigma}^{2}}{H^{2}}=-2a^{2}\frac{\dot{H}}{H^{2}}\equiv
2a^{2}\epsilon.
\end{equation}
We have introduced the slow-roll parameter $\epsilon$. The second order action
can be rewritten as%
\begin{equation}
S_{\left(  2\right)  }=\frac{1}{2}\int d\tau d^{3}x\left[  \left(  v^{\prime
}\right)  ^{2}-\left(  \partial_{i}v\right)  ^{2}+\frac{z^{\prime\prime}}%
{z}v^{2}\right]  .
\end{equation}

To quantize this action first define the canonical conjugate momentum of $v$,
and then impose the usual commutation relation%
\begin{equation}
\Pi_{v}=\frac{\partial L}{\partial v^{\prime}}=v^{\prime};\ \ \left[
v(\tau,\mathbf{x}),\Pi_{v}(\tau,\mathbf{x}^{\prime})\right]  =i\hbar
\delta^{\left(  3\right)  }(\mathbf{x}-\mathbf{x}^{\prime}).
\end{equation}
Henceforth we shall set $\hbar=1$.

Now one performs a plane-wave expansion of the now quantum operator $\hat
{v}(\tau,\vec{x})$ in Fourier space
\begin{equation}
\hat{v}(\tau,\vec{x})=\int\frac{d^{3}\mathbf{{k}}}{(2\pi)^{3}}[v_{k}(\tau
)\hat{a}_{\mathbf{k}}e^{i\mathbf{k}\cdot\mathbf{x}}+v_{k}^{\ast}(\tau)\hat
{a}_{\mathbf{k}}^{\dagger}e^{-i\mathbf{k}\cdot\mathbf{x}}]
\end{equation}
Requiring the canonical commutation relation between $\hat{a}_{\mathbf{k}%
}(\tau)$ and $\hat{a}_{\mathbf{k}}^{\dagger}(\tau)$, $[\hat{a}_{\mathbf{k}%
}(\tau),\hat{a}_{\mathbf{k}^{\prime}}^{\dagger}(\tau)]=(2\pi)^{3}\delta
^{(3)}(\mathbf{k}-\mathbf{k}^{\prime})$, we will obtain the Wronskian
condition for the mode function $v_{k}(\tau)$%

\begin{equation}
\left(  v_{k}^{\ast}v_{k}^{\prime}-v_{k}^{\prime\ast}v_{k}\right)  =-i\text{ }
\label{normalization}%
\end{equation}
The mode function in momentum space satisfies the Mukhanov-Sasaki equation
\begin{equation}
v_{k}^{\prime\prime}\left(  \tau\right)  +\left(  k^{2}-\frac{z^{\prime\prime
}}{z}\right)  v_{k}\left(  \tau\right)  =0\text{ } \label{Muk}%
\end{equation}

Upon introducing the second slow-roll parameter%
\begin{equation}
\eta=-\frac{\overset{\cdot\cdot}{\sigma}}{H\dot{\sigma}}%
\end{equation}
one can express $\frac{z^{\prime\prime}}{z}$ in terms of the first and second
slow-roll parameters%
\begin{equation}
\frac{z^{\prime\prime}}{z}=2a^{2}H^{2}\left(  1-\frac{3}{2}\eta+\epsilon
+\frac{1}{2}\eta^{2}-\frac{1}{2}\epsilon\eta+\frac{1}{2H}\dot{\epsilon}%
-\frac{1}{2H}\dot{\eta}\right)  \label{z''/z}%
\end{equation}

In general, Eq.$\left(  \ref{Muk}\right)  $ with $z^{\prime\prime}/z$ given in
Eq.$\left(  \ref{z''/z}\right)  $ is difficult to solve analytically. For a
special subset of general theories where $\epsilon$ and $\eta$ are
approximately constants, the equation is analytically solvable. In this
special case $z^{\prime\prime}/z$ can be written as%
\begin{equation}
\frac{z^{\prime\prime}}{z}=\frac{\nu^{2}-\frac{1}{4}}{\tau^{2}}\text{
\ \ where \ }\nu=\frac{1-\eta+\epsilon}{1-\epsilon}+\frac{1}{2}
\label{standard fit}%
\end{equation}
and the analytic solution for $v_{k}\left(  \tau\right)  $ is given in terms
of Bessel functions
\begin{equation}
v_{k}\left(  \tau\right)  =\alpha\sqrt{\tau}J_{\nu}\left(  k\tau\right)
+\beta\sqrt{\tau}Y_{\nu}\left(  k\tau\right)  \label{Bessel}%
\end{equation}

Where $\alpha$ and $\beta$ are two complex parameters. A well known example
for this special case is that of power law inflation, $a=c\tau^{p},$ where
$\epsilon=\eta=\frac{p+1}{p}$. One then obtains $z=\sqrt{\frac{2c^{2}(p+1)}%
{p}} \tau^{p}$, which gives $z^{\prime\prime2}$ or $\nu=-p+1/2,$ and the
comoving horizon ${(aH)}^{-1}=\tau/p$. Pure de Sitter expansion is the
specific case of $p=-1$. For genuine de Sitter inflation, $z$ vanishes, which
leads to an exactly scale-invariant power spectrum that is now observationally
disfavored \cite{Komatsu:2010}. This implies that inflation must deviate from
the pure de Sitter case.

The solutions to Eq.$(\ref{Muk})$ can be written either as linear combinations
of Bessel functions, $J_{n}(x)$ and $Y_{n}(x)$, or Hankel functions,
$H_{n}^{(1)}(x)$ and $H_{n}^{(2)}(x)$
\begin{align}
v_{k}\left(  \tau\right)   &  =\alpha\sqrt{\tau}J_{-p+\frac{1}{2}}\left(
k\tau\right)  +\beta\sqrt{\tau}Y_{-p+\frac{1}{2}}\left(  k\tau\right)
\label{v2}\\
&  =\widetilde{\alpha}\sqrt{\tau}H_{-p+\frac{1}{2}}^{\left(  1\right)
}\left(  k\tau\right)  +\widetilde{\beta}\sqrt{\tau}H_{-p+\frac{1}{2}%
}^{\left(  2\right)  }\left(  k\tau\right)
\end{align}
The Wronskian condition Eq.$(\ref{normalization})$ requires
\begin{equation}
\alpha^{\ast}\beta-\alpha\beta^{\ast}=-\frac{i\pi}{2}\text{ or }\left\vert
\widetilde{\alpha}\right\vert ^{2}-\left\vert \widetilde{\beta}\right\vert
^{2}=1\text{ } \label{normalize}%
\end{equation}
When the solution is expressed in terms of Hankel functions, there is a
natural place where the boundary condition may be imposed, that is when
$\left\vert k\tau\right\vert \gg1$, or equivalently when the comoving
wavelength is deep inside the comoving Hubble radius. The asymptotic forms of
the Hankel functions become positive and negative frequency modes
\begin{equation}
\lim_{\left\vert k\tau\right\vert \gg1}\sqrt{\tau}H_{-p+\frac{1}{2}}%
^{(2)}(k\tau)=e^{-ik\tau}\,\,;\,\,\lim_{\left\vert k\tau\right\vert \gg1}%
\sqrt{\tau}H_{-p+\frac{1}{2}}^{(1)}(k\tau)=e^{ik\tau}. \label{hankel}%
\end{equation}
The fact that in the far past the solution approaches those of Minkowski space
can be seen in the behavior of $z^{\prime\prime}/z$ displayed in
Fig.$(\ref{zppzfit})$

\begin{figure}[ptb]
\label{zppzfit} \includegraphics[scale = .5]{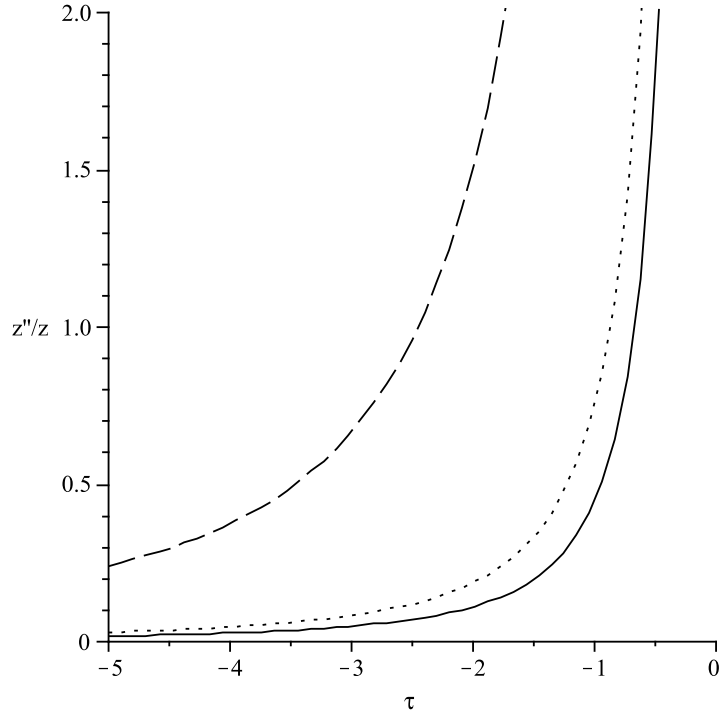}\caption{A plot of
$p(p-1)/\tau^{2}$ for the cases: $p=-2$ given by the dashed line, $p=-1/2$
given by the dotted line, and $p=-1/3$ given by the solid line.}%
\label{zppzfit}%
\end{figure}

The vanishing behavior of $z^{\prime\prime}/z$ in this asymptotic region
ensures that solutions to Eq.$(\ref{Muk})$ reduce to the Minkowski type in the
far past. Thus, for these classes of inflationary models there is a natural
boundary condition that the solution should approach the positive frequency
ougoing mode with no incoming modes
\begin{equation}
\lim_{\tau\rightarrow-\infty}v_{k}(\tau)=\frac{e^{-ik\tau}}{\sqrt{2k}}
\label{bc}%
\end{equation}
This form is seen to match that of $H_{p-\frac{1}{2}}^{(1)}(k\tau)$ in
Eq.(\ref{hankel}). The boundary condition Eq.(\ref{bc}) is known as the
Bunch-Davies vacuum \cite{BunchDavies}. This has the effect of setting
$\tilde{\alpha}=1$ and $\tilde{\beta}=0$ in Eq.(\ref{v2}). This appears as a
natural choice, as one may think intuitively that at the beginning of time,
all the particles (or positive frequency modes) should move forward in time,
thus eliminating the possibility of having a contribution from the
$H_{p-\frac{1}{2}}^{(2)}(k\tau)$ term.

We would like to stress that although $\tau\rightarrow-\infty$ is a legitimate
limit formally, but physically there will exist a time where the physical
wavelength will be comparable to the Planck length where quantum gravity
effects should take place. This means in that region, the background evolution
can no longer be treated classically. Due to the lack of a full quantum
gravity theory, the boundary condition may be imposed at some later time where
the physical wavelength is greater than the Planck length. The effect of
setting the boundary condition at a finite time may be that the state does not
reside in the ground state, but rather in some squeezed or distorted state
\cite{Danielsson trans Planck,New Physics CMB,Grishchuk:1993}.

For general energy contents of the universe, the form of the scale factor will
no longer be a simple power-law (although during times where a single
component is the dominant contributor to the stress-energy such as during
matter or radiation domination, the power-law form is a good approximation).
As an approximation, a standard analytic approach is to assume the expansion
is approximately de Sitter, $p\approx-1$, and therefore $\epsilon\approx0$ .
Together with the smallness condition of the second slow-roll parameter
$\eta=-\frac{\overset{\cdot\cdot}{\sigma}}{H\dot{\sigma}}\ll1$ using
Eq.$\left(  \ref{standard fit}\right)  $ one finds
\begin{equation}
\frac{z^{\prime\prime}}{z}\approx\frac{a^{\prime\prime}}{a}\approx\frac
{2}{\tau^{2}} \label{desitter}%
\end{equation}
Under this assumption, the solution of the Mukhanov-Sasaki equation is
\begin{equation}
v_{k}\left(  \tau\right)  =\tilde{\alpha}\sqrt{k\tau}H_{3/2}^{(1)}%
(k\tau)+\tilde{\beta}\sqrt{k\tau}H_{3/2}^{(2)}(k\tau)
\end{equation}
where $\tilde{\alpha}$ and $\tilde{\beta}$ are two complex parameters with
four degrees of freedom, one of which is fixed by Eq.(\ref{normalize}), two
more by Eq.(\ref{bc}), leaving one irrelevant phase undetermined. With these
conditions, the solution for the mode function is
\begin{equation}
v_{k}\left(  \tau\right)  =\frac{e^{-ik\tau}}{\sqrt{2k}}\left(  1-\frac
{i}{\kappa\tau}\right)  \label{BD mode function}%
\end{equation}
This leads to the well known relations for the scalar power-spectrum,
$P_{\mathcal{R}}$, the spectral index $n_{s}$, and the running of the spectral
index $\alpha_{s}$
\begin{align}
&  P_{\mathcal{R}}\equiv\left.  \frac{\left\vert v_{k}(\tau)\right\vert ^{2}%
}{z^{2}}\right\vert _{k=aH}=\left.  \frac{H^{2}}{4k^{3}\epsilon}\right\vert
_{k=aH}\label{scalar power}\\
&  n_{s}-1\equiv\left.  \frac{d\ln(k^{3}P_{\mathcal{R}})}{d\ln k}\right\vert
_{k=aH}=\left.  \frac{k}{k^{3}P_{\mathcal{R}}}\frac{d(H^{2}/\epsilon)}{d\tau
}\frac{d\tau}{dk}\right\vert _{\tau=\tau_{\ast}}\label{spectral}\\
&  \alpha_{s}\equiv\left.  \frac{dn_{s}}{d\ln k}\right\vert _{k=aH}
\label{running}%
\end{align}
Note that each relation is to be evaluated at horizon crossing, $k=aH$ (or
equivalently $\tau=\tau_{\ast}$), due to the fact that the perturbation is
frozen when the comoving wavelength becomes stretched outside the comoving
Hubble radius.

The accuracy of this program strongly depends on the quality of how well the
approximation of the curve $z^{\prime\prime}/z$ compares with the true
function $z^{\prime\prime}/z$ . Applying these equations to models where
$z^{\prime\prime}/z$ is not well fit by the power law and slow-roll
approximations may result in serious deviations from the observable
predictions. This is precisely the problem we will address in Section
\ref{vacuum}. Before doing so we will first give an overview of quantization
in the tensor sector.

\subsection{Tensor Perturbations}

The calculation for tensor perturbations mirrors that of the scalar
perturbation. The starting point is once again the perturbed FRW metric with a
transverse, traceless tensor perturbation, $h_{ij}$ (we have omitted the
scalar and vector perturbations seen previously which has no effect on the
tensor mode evolution due to decoupling of the scalar, vector, and tensor
sectors)
\begin{equation}
ds^{2}=a^{2}\left(  -d\tau^{2}+\left(  \delta_{ij}+h_{ij}(\mathbf{x}%
,\tau)\right)  dx^{i}dx^{j}\right)
\end{equation}

Next, one expands the Einstein-Hilbert action to second order in the
perturbation
\begin{equation}
S_{\left(  2\right)  }=\frac{M_{pl}^{2}}{8}\int d\tau dx^{3}a^{2}\left[
(h_{ij}^{\prime}(\mathbf{x},\tau))^{2}-(\partial_{l}h_{ij}(\mathbf{x}%
,\tau))^{2}\right]
\end{equation}
As in the scalar case, one expands $h_{ij}$ in Fourier space in terms of plane
waves with modes $h_{\mathbf{k}}^{s}$
\[
h_{ij}(\mathbf{x},\tau)=\int\frac{d^{3}k}{(2\pi)^{3}}\sum_{s=\times
,\,+}\epsilon_{ij}^{s}h_{\mathbf{k}}^{s}(\tau)e^{i\mathbf{k}\cdot\mathbf{x}}%
\]
where $\epsilon_{ij}^{s}$ are the spin-two polarization tensors. One can then
make the definition
\begin{equation}
\mu_{k}^{s}\left(  \tau\right)  =\frac{1}{2}ah_{k}^{s}\left(  \tau\right)
\end{equation}
which leads to the action (here we have set $M_{Pl}=1$)
\begin{equation}
S_{\left(  2\right)  }=\sum_{s}\frac{1}{2}\int d\tau d^{3}k\left[  \left(
\mu_{k}^{s\prime}\right)  ^{2}-\left(  k^{2}-\frac{a^{\prime\prime}}%
{a}\right)  \left(  \mu_{k}^{s}\right)  ^{2}\right]
\end{equation}
This action gives similar equations of motion as Eq.$\left(  \ref{Muk}\right)
$
\begin{equation}
\mu_{k}^{s\prime\prime}\left(  \tau\right)  +\left(  k^{2}-\frac
{a^{\prime\prime}}{a}\right)  \mu_{k}^{s}\left(  \tau\right)  =0\text{ }%
\end{equation}
For a scale factor of the power-law form, the calculation follows exactly as
in the scalar case while the fit function in terms of slow-roll parameter is
now%
\begin{equation}
\frac{a^{\prime\prime}}{a}=\frac{\mu^{2}-\frac{1}{4}}{\tau^{2}}\text{
\ \ where }\mu=\frac{1}{1-\epsilon}+\frac{1}{2}. \label{tensor fit}%
\end{equation}

In the case of quasi de Sitter inflation, $\epsilon\approx0$, the power
spectrum for a single polarization of the tensor modes toward the end of
inflation is
\begin{equation}
P_{t}=\left.  4\frac{\left\vert \mu_{k}\right\vert ^{2}}{a^{2}}\right\vert
_{k=aH}=\left.  \frac{2H^{2}}{k^{3}}\right\vert _{k=aH} \label{tensor power}%
\end{equation}
which differs from the form of the scalar result in that the slow-roll
parameter $\epsilon$ is absent in the denominator. The full power spectrum is
then twice this (due to two polarization states) $P_{h}=2P_{t}=4H^{2}/k^{3}$,
which leads to a small tensor to scalar ratio $r=P_{h}/P_{\mathcal{R}%
}=16\epsilon$ when the slow-roll parameter is small.

These results for the power spectra are obtained under the assumption that the
expansion is de Sitter or very nearly de Sitter in the sense that
Eq.(\ref{desitter}) is true. To obtain a more accurate predication one must
solve the Mukhanov equation on a model-by-model basis using the exact
$z^{\prime\prime}/z$ (or $a^{\prime\prime}/a$) numerically, along with
choosing a proper boundary condition accordingly. In the next section we will
institute such a procedure in order to quantize models where
Eq.$(\ref{desitter})$ is not a good approximation.

\section{Vacuum Selection}

\label{vacuumselection}

\label{vacuum}

The mode equation one needs to solve is
\begin{equation}
{v}_{k}^{\prime\prime2}-f(\tau))v_{k}=0 \label{master}%
\end{equation}
One can obtain the analytic expression for $f(\tau)$ from solving the
background equations. For the scalar case $f(\tau)$ is given by $z^{\prime
\prime}/z$, while for the tensor case it is $a^{\prime\prime}/a$. In the de
Sitter limit, its value is shown in Eq.(\ref{desitter}). In general, the
background evolution is not tractable analytically, and for the few cases
where the background solution is analytically solvable \cite{analytic
background,analytic Easther, analytic Barrow, analytic Barrow2, analytic Ellis
and Madsen,analytic Easther2,analytic Lidsey,analytic Mimoso}, the $\tau$
dependence in $f(\tau)$ may be so complicated that finding an analytic
solution for Eq.(\ref{master}) becomes impossible. It is worth mentioning that
the equation is equivalent to a time independent Schrodinger equation in one
dimension when $\tau$ is regarded as space coordinate and $f\left(
\tau\right)  $ is the potential for the wave function $v_{k}\left(
\tau\right)  .$ This analogy will become apparent when we introduce our method
of solving Eq.$\left(  \ref{master}\right)  $, which is nothing but the usual
WKB approximation in quantum mechanics.

Under the condition that the background evolution of the metric is known for a
particular inflationary model, one can determine whether the approximation
Eq.(\ref{desitter}) will be applicable for the model under consideration. If
so, then the system may be solved using the standard approach outlined in the
preceding section. However, this is not the case for a great number of
inflationary models. One can understand that this is so because the mechanism
used in stopping inflation may falsify the standard approximation.
Additionally, the initial conformal time can not always be pushed back to
negative infinity where one would impose the Bunch-Davies boundary condition
as is the case when the expansion is truly power-law.

The essential idea of our method is that, since the information one needs in
order to calculate observables to compare with experiment is the value of the
mode function at horizon crossing (which is much later than the asymptotic
time $\tau\rightarrow-\infty$), it may therefore be advantageous to place the
physical boundary condition closer to the time where we require accurate
information. Imposing the boundary condition at negative infinity may lead to
a situation where the approximate form for $f(\tau)$ in Eq.(\ref{desitter})
has deviated greatly from the actual solution due to the lengthy intervening
period of evolution. Although we are placing the boundary condition nearer the
era of observable inflation, we will continue to mimic the idea of BD vacuum
selection in that we look at the time $\tau=p$ where the wavelength of the
mode is deep inside the horizon, and the effect of cosmic expansion is
relatively small. The situation can then be approximated as physics in
Minkowski space.

In this section we will first demonstrate the method with an explicit example
before going on to comment on considerations on applying the method in general.

\subsection{A Specific Example}

The model we use as an example has background evolution that is analytically
solvable \cite{analytic background}. This particular model provides an
interesting picture where the Big Bang is connected to inflation with a
specific time delay. The behavior of $z^{\prime\prime}/z$ is very different
from the slow roll plus de Sitter limit, while the term $a^{\prime\prime}/a$
is asymptotically equal to the de Sitter limit. We will apply our method to
obtain the scalar power spectrum and constrain the tensor power spectrum from observation.

Our point here is to show that there exist examples such that $z^{\prime
\prime}/z$ can not be approximated by the standard fit function, thus there is
a \emph{need} of introducing new method of solving Mukhanov-Sasaki equation.
We would like to point out this model does not provide a mechanism of stopping
inflation, therefore, even though our method predicts the correct power
spectrum in a certain parameter space, the phenomenologically viable parameter
space is expected to change when the stopping mechanism is introduced.
Therefore the model in question may be viewed as a demonstration tool (due to
the attractive property that it is analytically solvable at the background
level) rather than a fully complete model.

We begin with an action of the form Eq.$\left(  \ref{action}\right)  $ with
the scalar potential
\begin{equation}
V\left(  \sigma\right)  =\left(  \frac{6}{\kappa^{2}}\right)  ^{2}\left(
c\sinh^{4}\left(  \sqrt{\frac{\kappa^{2}}{6}}\sigma\right)  +b\cosh^{4}\left(
\sqrt{\frac{\kappa^{2}}{6}}\sigma\right)  \right)  \label{inflaton potential}%
\end{equation}
The potential contains the dimensionless free parameters $b$ and $c$, and
$\kappa^{-1}$ is the reduced Planck mass as before. The background solution
for all possible combinations of $c$ and $b$ has been classified in
\cite{analytic background}. We will use the case $c=64b>0$ as an example to
illustrate our method.

The behavior of the scale factor of this model is to initiate expansion at
$\tau=\tau_{BB}\approx0.92$ where the scale factor is exactly zero. Beginning
at a later time, $\tau_{I}$ $\approx2.87$, there is an inflationary period,
and finally, when $\tau$ approaches $\tau_{\infty}\approx7.4$, the scale
factor diverges. At this point the physical time, $t=\int a\left(
\tau\right)  d\tau$, will also diverge. Of course the finite value
$\widetilde{\tau}_{BB}\approx0.92$ is not physically significant since $\tau$
can be translated by an arbitrary amount. The solutions for the background
evolution of the scale factor $a\left(  \tau\right)  $ and inflaton
$\sigma(\tau)$ are as follows%
\begin{equation}
a\left(  \tau\right)  =\sqrt{\frac{1}{12}}\kappa\left(  \frac{E}{b}\right)
^{\frac{1}{4}}\left\{  2\left[  \frac{1-cn\left(  \frac{1}{2}\widetilde{\tau
}\right)  }{1+cn\left(  \frac{1}{2}\widetilde{\tau}\right)  }\right]
-\frac{1}{4}\left[  cn\left(  \widetilde{\tau}\right)  \right]  ^{2}\right\}
^{\frac{1}{2}}\label{AE}%
\end{equation}
and%
\begin{equation}
\sigma\left(  \tau\right)  =\frac{1}{\kappa}\sqrt{\frac{3}{2}}\ln\left(
\frac{1+\frac{1}{2\sqrt{2}}cn\left(  \widetilde{\tau}\right)  \left[
\frac{1+cn\left(  \frac{1}{2}\widetilde{\tau}\right)  }{1-cn\left(  \frac
{1}{2}\widetilde{\tau}\right)  }\right]  ^{\frac{1}{2}}}{1-\frac{1}{2\sqrt{2}%
}cn\left(  \widetilde{\tau}\right)  \left[  \frac{1+cn\left(  \frac{1}%
{2}\widetilde{\tau}\right)  }{1-cn\left(  \frac{1}{2}\widetilde{\tau}\right)
}\right]  ^{\frac{1}{2}}}\right)  \label{sigma as function of z and t}%
\end{equation}
where $\widetilde{\tau}\equiv2\left\vert cE\right\vert ^{\frac{1}{4}}\tau$.
$E$ is a free parameter of this model which determines the scale of $a$,
therefore $E$ can be chosen so that $a$ is normalized in the conventional way,
$a\left(  today\right)  =1$. In the present application we will choose $E$ so
that
\begin{equation}
2\left\vert cE\right\vert ^{\frac{1}{4}}=1\text{ \ or \ }\widetilde{\tau
}\equiv\tau.
\end{equation}

\begin{figure}[ptb]
\includegraphics[scale=.5]{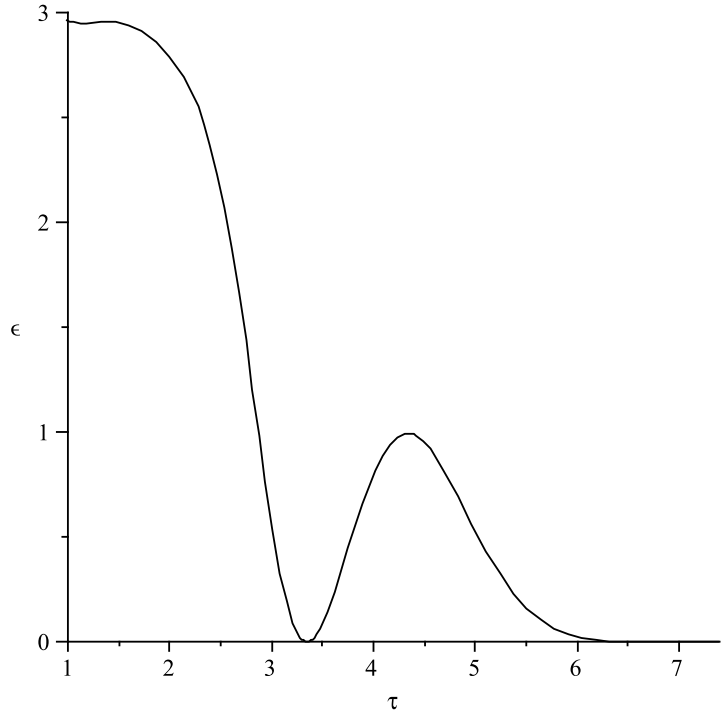}\caption{The slow-roll parameter
$\epsilon$ as a function of conformal time.}%
\label{epsilon}%
\end{figure}

\begin{figure}[ptb]
\includegraphics[scale=.5]{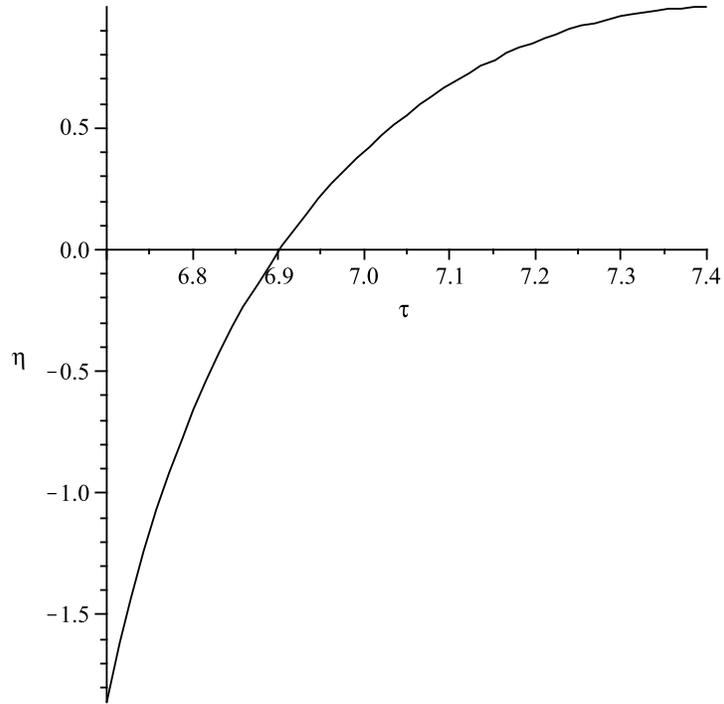}\caption{The slow-roll parameter
$\eta$ as a function of conformal time.}%
\label{eta}%
\end{figure}

The term $cn\left(  \frac{1}{2}\widetilde{\tau}\right)  \equiv cn\left(
\frac{1}{2}\widetilde{\tau}|\frac{1}{2}\right)  $ is the Jacobi elliptic
function\cite{abramcwtz}. Following the standard method, one needs to
determine whether $\epsilon$ and $\eta$ are approximately constants. If so,
following Eqs.$\left(  \ref{standard fit}\right)  $ and $\left(
\ref{tensor fit} \right)  $, one can determine the fit function of
$z^{\prime\prime}/z$ and $a^{\prime\prime}/a$. Since the analytic solution of
this model is known, we can plot the exact curves for $\epsilon$ and $\eta$ as
in Fig.\ref{epsilon} and Fig.\ref{eta}).

One can easily see that the asymptotic values of $\epsilon$ and $\eta$
approach the constant values $0$ and $1$ respectively. These values will
result in the breakdown of Eq.(\ref{standard fit}) for $z^{\prime\prime}/z$
and $2/(\tau-\tau_{\infty})^{2}$ for $a^{\prime\prime}/a$. From the above
expression we can also compute $z^{\prime\prime}/z$ and $a^{\prime\prime}/a$.
The result is shown in the following plots along with the fitting function

\begin{figure}[ptb]
\includegraphics[scale = .5]{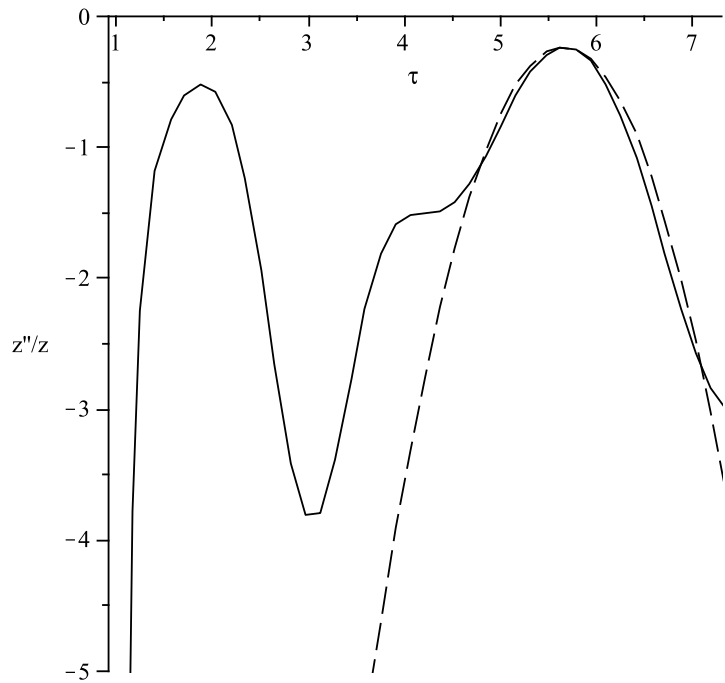}\caption{A plot of the function
$z^{\prime\prime}/z$ as a function of conformal time. The solid line is the
actual value of the function $z^{\prime\prime}/z$, and the dashed line gives
the fit function.}%
\label{zppz}%
\end{figure}

\begin{figure}[ptb]
\includegraphics[scale = .5]{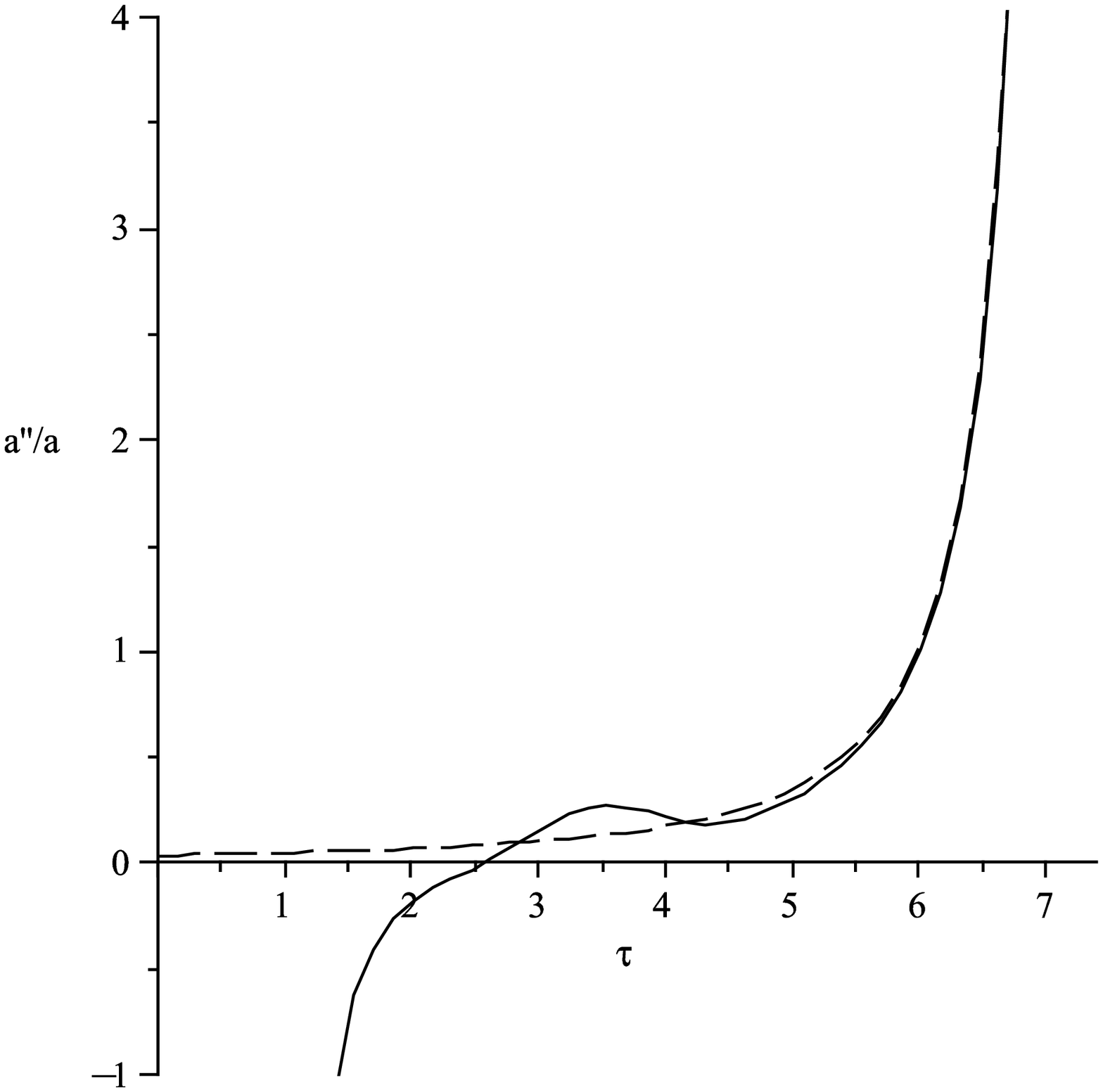}\caption{A plot of the function
$a^{\prime\prime}/a$ as a function of conformal time. The solid line is the
actual value of the function $a^{\prime\prime}/a$, and the dashed line gives
the fit function.}%
\label{appa}%
\end{figure}

The fit function plotted in Fig.$(5)$ is the predicted fit function for
$a^{\prime\prime}/a$ which is \newline$2/(\tau-7.4)^{2}$, while the fit
function for Fig.$\left(  4\right)  $ is a quadratic function
\begin{equation}
f_{fit}(\tau)=m(\tau-p)^{2}+h
\end{equation}
with parameters
\begin{equation}
m=-1.2,p=5.66,h=-0.23
\end{equation}

In the case of the scalar perturbation, since the formula for the standard fit
function breaks down, we now introduce a new quadratic fit function according
to the following rules. The parameter $p$ is designed to match the point where
$f(\tau)$ has slope zero while $h$ is designed to match the value of $f\left(
\tau\right)  $ at $\tau=p$. The value of $m$ is the only parameter in the fit
function, but it has to be adjusted such that the curve of $f_{fit}(\tau)$
approaches the actual $f\left(  \tau\right)  $ in the region of physical
interest, namely the region where the mode exits the horizon. That is to say
we want a better fit for $\tau>p$.  The region $\tau<p$ corresponds to the time before the beginning of observable inflation, or before about 60 e-folds
before the end of inflation.  The perturbation generated during the preceding
periods have yet to re-enter our horizon, thus they have produced no observable effect.
(Since the universe is currently accelerating again, it is not clear to what extent regions corresponding to time intervals before the last 60 e-folds will contribute to observable effects in the
future.)

Once the fit function is chosen, one can then proceed to solve the
Mukhanov-Sasaki equation with the new fit function. In the current example,
the two independent solutions to Eq.$\left(  \ref{master}\right)  $ with
$f\left(  \tau\right)  =f_{fit}\left(  \tau\right)  $ are the hypergeometric
functions $S_{k1}$ and $S_{k2}$. The general mode function is thus%

\[
v_{k}\left(  \tau\right)  =\alpha S_{k1}\left(  \tau\right)  +\beta
S_{k2}\left(  \tau\right)
\]
where $\alpha$ and $\beta$ are two complex parameters that can be parametrized
by four real parameters%
\begin{equation}
\alpha=r_{1}e^{i\theta_{1}},\beta=r_{2}e^{i\theta_{2}}%
\end{equation}
written in the real parameters
\begin{equation}
v_{k}\left(  \tau\right)  =e^{i\theta_{1}}\left(  r_{1}S_{1}+r_{2}e^{i\left(
\theta_{2}-\theta_{1}\right)  }S_{2}\right)
\end{equation}
$r_{1},r_{2},\theta_{1},\theta_{2}$ are to be determined by the boundary
conditions. It can be easily seen that when the quantity of interest is the
expectation value of $v_{k}\left(  \tau\right)  $ the parameter $\theta_{1}$
is an irrelevant phase.

The exact solutions $S_{k1}$ and $S_{k2}$ are%
\begin{align}
S_{k1}\left(  \tau\right)   &  =e^{\left(  -\frac{\tau}{2}+p\right)  \sqrt
{m}\tau}hypergeom\left(  \frac{-1}{4}\frac{-\sqrt{m}+k^{2}-h}{\sqrt{m}}%
,\frac{1}{2},\sqrt{m}\left(  \tau-p\right)  ^{2}\right) \\
S_{k2}\left(  \tau\right)   &  =\left(  \tau-p\right)  e^{\left(  -\frac{\tau
}{2}+p\right)  \sqrt{m}\tau}hypergeom\left(  \frac{-1}{4}\frac{-3\sqrt
{m}+k^{2}-h}{\sqrt{m}},\frac{3}{2},\sqrt{m}\left(  \tau-p\right)  ^{2}\right)
\end{align}
where the hypergeometric function is defined as\cite{abramcwtz}%
\begin{equation}
hypergeom\left(  a,b,z\right)  \equiv1+\frac{a}{b}z+\frac{a\left(  a+1\right)
}{b\left(  b+1\right)  2!}z^{2}+...
\end{equation}
The boundary condition Eq.$\left(  \ref{normalization}\right)  $ will require
$\alpha$ and $\beta$ to satisfy%

\[
1=i\left(  v_{k}^{\ast}v_{k}^{\prime}-v_{k}^{\prime\ast}v_{k}\right)  =%
\begin{array}
[c]{c}%
i\left(  \alpha^{\ast}\beta-\alpha\beta^{\ast}\right)  \text{ if }m<0\\
i\left(  \alpha^{\ast}\beta-\alpha\beta^{\ast}\right)  e^{\sqrt{m}p^{2}}\text{
if }m>0
\end{array}
\]

or%
\begin{equation}
\alpha^{\ast}\beta-\alpha\beta^{\ast}=2ir_{1}r_{2}\sin\left(  \theta
_{2}-\theta_{1}\right)  =-iC\left(  m,p\right)
\end{equation}

Where $C$ is a number which depends on the value of $m$ and $p$. When $m<0,$
$C=1$, when $m>0,$ $C=e^{-\sqrt{m}p^{2}}.$ In the specific example
discussed in this paper $m=-1.2<0$ For completeness purposes, we have
also included the expression for $C$ for postive $m.$

The next step is to choose a corresponding boundary condition. Since Fig.(4)
shows $z^{\prime\prime}/z<0$ in the whole region of physical conformal time,
$0.92<\tau<7.4$, the solution to Eq.$\left(  \ref{master}\right)  $ will be a
combination of oscillating waves. We introduce the WKB type solution
\begin{equation}
v_{k}\left(  \tau\right)  =A\left(  \tau\right)  e^{i\phi\left(  \tau\right)
}%
\end{equation}
where $A\left(  \tau\right)  $ and $\phi\left(  \tau\right)  $ are two real
functions. Inserting this ansatz into Eq.$\left(  \ref{master}\right)  $ one
obtains
\begin{equation}
A^{\prime\prime}+2iA^{\prime}\phi^{\prime}+iA\phi^{\prime\prime}-A\left(
\phi^{\prime}\right)  ^{2}=-\omega^{2}A
\end{equation}
where $\omega\left(  \tau\right)  =\sqrt{k^{2}-f\left(  \tau\right)  }$. This
complex equation can be separated into two real equations
\begin{align}
\label{real1}A^{\prime\prime}-A\left(  \phi^{\prime}\right)  ^{2}  &
=-\omega^{2}A\\\label{real2}
\left(  A^{2}\phi^{\prime}\right)  ^{\prime}  &  =0
\end{align}
The second equation can be solved easily
\begin{equation}
A=\frac{\pm q}{\sqrt{\phi^{\prime}}}%
\end{equation}
where $q$ is an arbitrary constant.

To solve Eq.(\ref{real1}) we invoke the WKB approximation which assume $A$ is
slowly varying with conformal time, and then the second derivative term can be
neglected. This approximation is true when the potential, $f(\tau)$, is slowly
varying with conformal time. This is true when the function $f\left(
\tau\right)  $ is at a local extrema, which corresponds to the very moment
when we impose the boundary condition. Under this approximation the
Eq.(\ref{real1}) can be solved analytically
\begin{equation}
\phi\left(  \tau\right)  =\pm\int\omega\left(  \tau\right)  d\tau
\end{equation}
The full solution can than be expressed as a linear combination of two modes%
\begin{equation}
v_{k}\left(  \tau\right)  =c_{1}\frac{e^{i\int\omega\left(  \tau\right)
d\tau}}{\sqrt{\omega}}+c_{2}\frac{e^{-i\int\omega\left(  \tau\right)  d\tau}%
}{\sqrt{\omega}}%
\end{equation}
In the vicinity of $\tau=p,$ $\omega\left(  \tau\right)  $ is a constant,
$\omega_{\ast}=\sqrt{k^{2}-h}.$ The boundary condition we impose is to choose
the parameter $\alpha$ and $\beta$ so that in the vicinity of
$\tau\approx p$, the solution is a linear combination of incoming and outgoing
plane waves satisfying the usual Klein-Gordon normalization
\begin{equation}
\text{when }\tau\approx p,\text{ }v_{k}\left(  \tau\right)  \approx
a\frac{e^{-i\omega_{\ast}\tau}}{\sqrt{2\omega_{\ast}}}+b\frac{e^{i\omega
_{\ast}\tau}}{\sqrt{2\omega_{\ast}}}\text{ \ where, }\left\vert a\right\vert
^{2}-\left\vert b\right\vert ^{2}=1\label{new BC}%
\end{equation}
The proposed boundary condition does invoke more parameters than the
Bunch-Davies boundary condition since we are not setting $a$ to 1 and $b$ to 0, but instead keeping them as general parameters.  As will be shown later, the
phenomenologically viable parameter space requires $a$ close to one and $b$ close to zero. This means the boundary condition is close to a purely
outgoing wave with smal amount of incoming wave, and thus close to the standard BD vacuum. It is
interesting that a small amount of incoming waves is necessary to produce
correct predictions. Unlike the Bunch-Davies boundary condition, which is
imposed at the beginning of time, our method imposes the boundary condition at
a finite time, $\tau=p$, where there is no reason to assume there only exists
an outgoing mode. One may criticize that our method introduces too many new
parameters in the boundary condition. However, this may also be the case if
$z^{\prime\prime}/z$ can be approximated by Eq.$\left(  \ref{standard fit}%
\right)  $ where the standard Bunch-Davies condition is meaningful. This is
because transplanckian physics could significantly alter the initial condition
such that the Bunch-Davies condition is not valid at all \cite{Danielsson
trans Planck}. Thus, although more parameters are introduced in our method,
the freedom is not necessarily more than the standard method.

Now we express $a$ and $b$ in terms of real parameters $a=a_{0}e^{i\theta_{a}%
},b=b_{0}e^{i\theta_{b}},$ $\left\vert a\right\vert ^{2}-\left\vert
b\right\vert ^{2}=1.$ According to the proposed boundary condition, the mode
function and its derivative are given by
\begin{align}
\left\vert v_{k}\left(  p\right)  \right\vert ^{2} &  =\left\vert
a\frac{e^{-i\omega p}}{\sqrt{2\omega}}+b\frac{e^{i\omega p}}{\sqrt{2\omega}%
}\right\vert ^{2}=\frac{1}{2\omega}+\frac{b_{0}^{2}+b_{0}\sqrt{1+b_{0}^{2}%
}\cos\left(  \tilde{\Delta}\right)  }{\omega},\\
\left\vert v_{k}^{\prime}\left(  p\right)  \right\vert ^{2} &  =\left\vert
\frac{-i\omega ae^{-i\omega p}}{\sqrt{2\omega}}+\frac{i\omega be^{i\omega p}%
}{\sqrt{2\omega}}\right\vert ^{2}=\frac{\omega}{2}+\omega\left(  b_{0}%
^{2}-b_{0}\sqrt{1+b_{0}^{2}}\cos\left(  \tilde{\Delta}\right)  \right)
\label{bc2}%
\end{align}
where $\tilde{\Delta}=2\omega p-\theta_{a}+\theta_{b}$. Here we have assumed
$\omega=\sqrt{k^{2}-h}>0$.  In the example we are discussing $h=-0.23$,
thus this assumption is always true for this case.  However, with a different fit function, it is conceivable that 
$h$ could be positive, giving the condition $k^{2} > h$ in such an instance.  The idea behind this choice is that at $\tau=p$, the equation becomes a
harmonic oscillator with constant frequency $\omega$. For a small period of
time near $p$, the solution should then approach the usual harmonic oscillator
where the vacuum is given by the lowest energy state. From Eq.$(\ref{bc2})$ we
find two more constraints that can be used to solve $r_{1}$, $r_{2}$, and
$(\theta_{2}-\theta_{1})$
\begin{align}
\left\vert v_{k}\left(  p\right)  \right\vert ^{2} &  =\left\vert
\alpha\right\vert ^{2}\left\vert e^{\frac{p^{2}}{2}\sqrt{m}}\right\vert
^{2}=\frac{1}{2\omega}+\frac{b_{0}^{2}+b_{0}\sqrt{1+b_{0}^{2}}\cos\left(
\tilde{\Delta}\right)  }{\omega}\\
\left\vert v_{k}^{\prime}\left(  p\right)  \right\vert ^{2} &  =\left\vert
\beta\right\vert ^{2}\left\vert e^{\frac{p^{2}}{2}\sqrt{m}}\right\vert
^{2}=\frac{\omega}{2}+\omega\left(  b_{0}^{2}-b_{0}\sqrt{1+b_{0}^{2}}%
\cos\left(  \tilde{\Delta}\right)  \right)
\end{align}

The solutions can be categorized according to the sign of $m$. We find that if
$m<0$
\begin{align}
r_{1}  &  =\sqrt{\frac{1}{2\omega}+\frac{b_{0}^{2}+b_{0}\sqrt{1+b_{0}^{2}}%
\cos\left(  \tilde{\Delta}\right)  }{\omega}},\\
r_{2}  &  =\sqrt{\frac{\omega}{2}+\omega\left(  b_{0}^{2}-b_{0}\sqrt
{1+b_{0}^{2}}\cos\left(  \tilde{\Delta}\right)  \right)  },\\
\left(  \theta_{2}-\theta_{1}\right)   &  =-\sin^{-1}\left(  \frac{1}%
{\sqrt{1+4\left(  b_{0}^{2}+b_{0}^{4}\right)  \sin^{2}\tilde{\Delta}}}\right)
\end{align}
while for $m>0$ we obtain
\begin{align}
r_{1}  &  =e^{-\frac{\sqrt{m}}{2}p^{2}}\sqrt{\left(  \frac{1}{2\omega}%
+\frac{b_{0}^{2}+b_{0}\sqrt{1+b_{0}^{2}}\cos\left(  \tilde{\Delta}\right)
}{\omega}\right)  }\\
r_{2}  &  =e^{-\frac{\sqrt{m}}{2}p^{2}}\sqrt{\left(  \frac{\omega}{2}%
+\omega\left(  b_{0}^{2}-b_{0}\sqrt{1+b_{0}^{2}}\cos\left(  \tilde{\Delta
}\right)  \right)  \right)  }\\
\left(  \theta_{2}-\theta_{1}\right)   &  =-\sin^{-1}\left(  \frac{1}%
{\sqrt{1+4\left(  b_{0}^{2}+b_{0}^{4}\right)  \sin^{2}\tilde{\Delta}}}\right)
\end{align}
Then for the mode functions we have for $m<0$
\begin{equation}
v_{k}\left(  \tau\right)  =e^{i\theta_{1}}\left(
\begin{array}
[c]{c}%
\sqrt{\frac{1}{2\omega}+\frac{b_{0}^{2}+b_{0}\sqrt{1+b_{0}^{2}}\cos\left(
\tilde{\Delta}\right)  }{\omega}}S_{1}\\
+e^{-i\sin^{-1}\left(  \frac{1}{\sqrt{1+4\left(  b_{0}^{2}+b_{0}^{4}\right)
\sin^{2}\tilde{\Delta}}}\right)  }\sqrt{\frac{\omega}{2}+\omega\left(
b_{0}^{2}-b_{0}\sqrt{1+b_{0}^{2}}\cos\left(  \tilde{\Delta}\right)  \right)
}S_{2}%
\end{array}
\right)
\end{equation}
and for $m>0$
\begin{equation}
v_{k}\left(  \tau\right)  =e^{i\theta_{1}}\left(
\begin{array}
[c]{c}%
e^{-\frac{\sqrt{m}}{2}p^{2}}\sqrt{\frac{1}{2\omega}+\frac{b_{0}^{2}+b_{0}%
\sqrt{1+b_{0}^{2}}\cos\left(  \tilde{\Delta}\right)  }{\omega}}S_{1}\\
+e^{-i\sin^{-1}\left(  \frac{1}{\sqrt{1+4\left(  b_{0}^{2}+b_{0}^{4}\right)
\sin^{2}\tilde{\Delta}}}\right)  }e^{-\frac{\sqrt{m}}{2}p^{2}}\sqrt
{\frac{\omega}{2}+\omega\left(  b_{0}^{2}-b_{0}\sqrt{1+b_{0}^{2}}\cos\left(
\tilde{\Delta}\right)  \right)  }S_{2}%
\end{array}
\right)
\end{equation}

With the above expressions, the mode function is uniquely determined (up to an
irrelevant phase $\theta_{1}$) by the fit function $f_{fit}\left(
\tau\right)  $ along with the parameters $b_{0}$ and $\left(  \theta
_{a}-\theta_{b}\right)  $ from the boundary condition. As usual, the mode
function will be evaluated at horizon crossing, where $k=aH$.

In order to produce an observationally consistent spectral index and its
running, we find the parameters $\tilde{\Delta}$ and $b_{0}$ are 0 and 0.075
respectively . The mode function then satisfies the boundary condition%
\begin{equation}
\text{at }\tau\approx p,v\text{ }_{k}\left(  \tau\right)  \approx
1.0028\frac{e^{-i\omega\tau}}{\sqrt{2\omega}}+0.075\frac{e^{i\omega\tau}%
}{\sqrt{2\omega}}\label{scalar mode}%
\end{equation}
One can derive the spectral index as well as the running of the spectral index
using definition Eq.$\left(  \ref{spectral}\right)  \left(  \ref{running}%
\right)  $ as function of conformal time

\begin{figure}[ptb]
\includegraphics{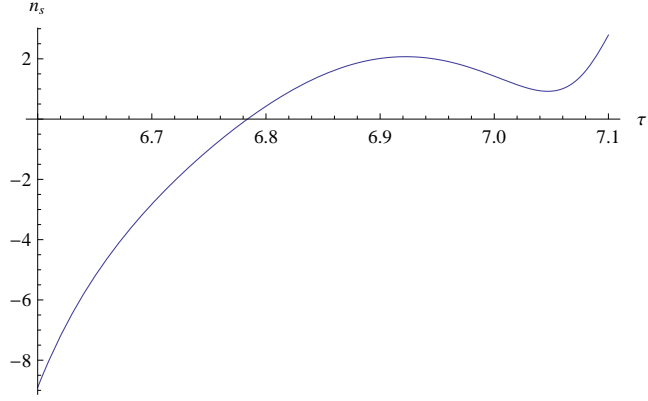}\caption{The scalar spectral index $n_{s}$ as a
function of conformal time.}%
\label{spectral}%
\end{figure}

\begin{figure}[ptb]
\includegraphics{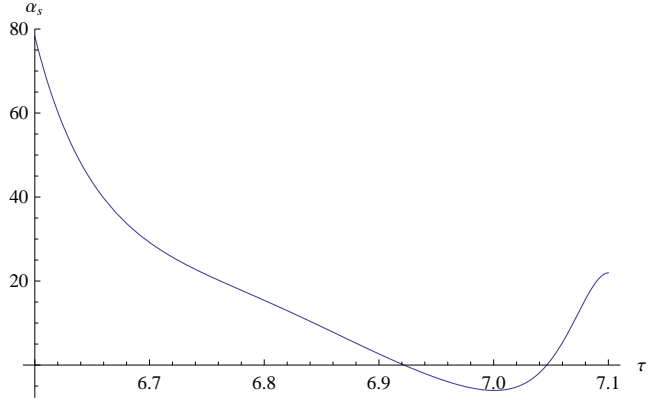}\caption{The running of the scalar spectral index
$\alpha_{s}$ as a function of conformal time.}%
\label{running}%
\end{figure}

At $\tau\approx7.05$ the spectral index is $n\approx0.95$, and the running of
the spectral index is $\alpha\approx0.$ This is within the allowed value of
WMAP constraints \cite{Komatsu:2010}.

Now, consider the tensor perturbation. The mismatch of the standard fit
function is shown in Fig.$\left(  5\right)  $. The fit function $2/(\tau
-7.4)^{2}$ is a good approximation in the asymptotic region. This is because
the first slow-roll parameter is approaching zero when $\tau>6.$ Despite the
success in the asymptotic region, the validity of Bunch-Davies boundary
condition is questionable. The reason is that Bunch-Davies vacuum imposes a
condition at a fictitious conformal time $\tau\rightarrow-\infty$, while in
the physical universe actually starts its expansion at a finite conformal time
$\tau\approx0.92.$ The limit $\tau\rightarrow-\infty$ is inapplicable in our
model thus the mode function evaluated from the evolution of such ,an ill
defined boundary condition is not reliable. Thus, we shall keep the standard
fit function but abandon the boundary condition that is imposed at an
unphysical time. We will proceed by obtaining a general solution of the mode
function, and then subsequently use the observational constraints to restrict
the possible parameters. Since the power spectrum of tensor perturbations has
yet to be observed, experimentally what we have to constrain is the ratio of
tensor perturbation to the scalar perturbation.

The most general mode function for the tensor modes is
\begin{align}
\mu_{k}\left(  \tau\right)   &  =a\sqrt{k\tau}H_{3/2}^{(1)}(k\tau
)+b\sqrt{k\tau}H_{3/2}^{(2)}(k\tau)\\
&  =a\frac{e^{-ik\tau}}{\sqrt{2k}}\left(  1-\frac{i}{k\tau}\right)
+b\frac{e^{ik\tau}}{\sqrt{2k}}\left(  1+\frac{i}{k\tau}\right)
\end{align}
From the definition $\left(  \ref{tensor power}\right)  $ one obtains
\begin{equation}
P_{h}=4\frac{H^{2}}{k^{3}}+8\frac{H^{2}}{k^{3}}\left(  b_{0}^{2}-b_{0}%
\sqrt{1+b_{0}^{2}}\cos\left(  \tilde{\bigtriangleup}\right)  \right)
\end{equation}
Together with P$_{\mathcal{R}}$ computed from Eq.$\left(  \ref{scalar mode}%
\right)  $ as well as the observational restriction on $r$, one can deduce an
upper bound for $\left(  2b_{0}^{2}-2b_{0}\sqrt{1+b_{0}^{2}}\cos\left(
\tilde{\bigtriangleup}\right)  \right)  .$ From the WMAP7 data
\cite{Komatsu:2010} $r<0.2$, which corresponds to $\left(  2b_{0}^{2}%
-2b_{0}\sqrt{1+b_{0}^{2}}\cos\left(  \tilde{\bigtriangleup}\right)  \right)
<0.16.$

\subsection{The General Case}

We would like to comment on how to generalize this program, such that it may
be implemented for a given background evolution of an inflationary model.
There is no reason, \emph{a priori}, that either Eq.$\left(
\ref{standard fit} \right)  $ in the standard method, or a quadratic function,
as introduced in the previous section, should be a good fit for $z^{\prime
\prime}/z$ in general.

Our proposal in this paper has been to point out that one should not apply the
standard fit function to all inflationary models without carefully examining
whether it is actually a good fit. If the analytic solution for the
Mukahanov-Sasaki equation is not attainable, one may wish to find a fit
function that is analytically solvable. There exist a number of analytically
solvable fit functions for $z^{\prime\prime}/z$, the quadratic function
introduced in the previous section is just one simple example.

Another more sophisticated example is the quartic function which renders
Eq.$\left(  \ref{Muk}\right)  $ solvable by linear combinations of the Heun
triconfluent function. Unfortunately in practice, the introduction of a more
complicated fitting function, such as the quartic function, may reduce the
predictivity of the model as there will tend to be more parameters in the fit
function that require matching rules to pin down their values. However the
rule of thumb is simple, the employed fit function should resemble the actual
curve at the region of interest, which is the moment of horizon crossing.
After that, one should impose the boundary condition at a meaningful time. As
we have demonstrated, using the WKB approximation at the moment when
$z^{\prime\prime}/z$ is flat is one sensible choice. Finally, one can use the
current observational data for the tensor to scalar ratio to constrain tensor
perturbation. However, since the power spectrum of the tensor mode has not
been observed, the constraints from the tensor sector are not overly restrictive.

\section{Conclusions}

\label{conclusions}

In order to make predictions testable by observations, inflation needs not
only a model, but suitable boundary conditions. Some models of inflation do
not seem to fall within the realm where the standard boundary conditions may
be naturally applied. With this in mind, we have discussed the introduction of
an alternative method which generalizes the standard approach of computing the
scalar and tensor power spectra. It is suggests that for those models whose
background is analytically solvable, one should re-examine their power
spectrum using our method and find how does their spectral index compare with
the results from standard method.

In general, this procedure will introduce additional parameters into the
model, thus allowing more accurate phenomenology, with the usual drawback that
introduction of more parameters decreases predictivity. This new method is
implemented on a model-by-model basis, hence the generic effects of this
approach have yet to be determined. For the specific example discussed in this
paper, we explored an inflationary model which has analytically solvable
background dynamics. We introduced a quadratic fit (of course, other models
may require more complicated fitting functions)for the function $z^{\prime
\prime}/z$ which appears in the Mukhanov-Sasaki equation for the scalar modes,
and imposed boundary conditions at finite conformal time, $\tau$. It was found
that near $\tau=7.05$ the spectral index and its running, both fall into a
phenomenologically acceptable range. This calculation gives an example for the
implementation of our approach, although the model in question is not fully
realized in the sense that it lacks a proper accounting for the cessation of
inflation in order to produce the requisite amount of e-fold expansion.

The capacity for altering the calculation, and thus the values, of observables
predicted by inflation via this new approach is clear. It may therefore be
possible that models which were hitherto discarded may need to be
re-investigated in the framework of this method.

\bigskip

\begin{acknowledgments}
We thank Itzhak Bars, Yi-Fu Cai, Yi-Zen Chu, and Tanmay Vachaspati for helpful discussions. The work of SHC was partially supported
by the US Department of Energy, grant number DE-FG03-84ER40168.  The work of SCH and JBD is supported in part by the Arizona State University Cosmology Initiative.
\end{acknowledgments}

\end{document}